\documentclass{elsarticle}
\usepackage{graphicx}%
\usepackage{multirow}%
\usepackage{amsmath,amssymb,amsfonts}%
\usepackage{amsthm}%
\usepackage{mathrsfs}%
\usepackage[title]{appendix}%
\usepackage{xcolor}%
\usepackage{textcomp}%
\usepackage{manyfoot}%
\usepackage{booktabs}%
\usepackage{algorithm}%
\usepackage{algorithmicx}%
\usepackage{algpseudocode}%
\usepackage{listings}%
\usepackage{verbatim}
\usepackage{enumitem}
 \usepackage{comment}
 \usepackage{amsmath}
 
 \journal{}
\begin{document}

\begin{frontmatter}

\title{Negative partisanship drives anti-frontrunner volatility in election dynamics}

\author[mymainaddress]{Jos\'e F.  Fontanari}

\address[mymainaddress]{Instituto de F\'{\i}sica de S\~ao Carlos, Universidade de S\~ao Paulo, 13566-590 S\~ao Carlos, S\~ao Paulo, Brazil}

\begin{abstract}
Negative partisanship---where voter mobilization is driven primarily by aversion toward an opposing candidate rather than positive alignment---profoundly reshapes modern electoral dynamics.  Here, we formulate a two-dimensional piecewise map to model campaign opinion dynamics, where voter abstention arises endogenously from a latent spectrum of voter activation barriers. We show that cross-candidate rejection sensitivity triggers a non-smooth collapse of abstention, dividing the phase space into an under-mobilized regime and a fully mobilized campaign regime. In symmetric elections with radicalized voter bases, spontaneous symmetry breaking via a pitchfork bifurcation gives rise to asymmetric polling attractors partitioned by mirror-symmetric basins of attraction, where early public perceptions---encoded in initial campaign conditions---act as a deterministic catalyst selecting candidate lead lock-in long before election day. For elevated cross-rejection, the symmetric equilibrium destabilizes via a period-doubling flip bifurcation, driving persistent anti-frontrunner polling swings. Rather than predicting static candidate hegemony, our model reveals that high negative partisanship turns elections into time-sensitive dynamical systems, where final election outcomes reflect discrete sampling snapshots of an oscillating campaign trajectory.
\end{abstract}

\end{frontmatter}

\section{Introduction}

Threshold models of collective behavior, pioneered by Granovetter~\cite{granovetter1978,granovetter1983,Fontanari_2026}, provide a mathematical baseline for understanding self-reinforcing social cascades and opinion formation~\cite{bikhchandani1992,watts2002,jusup2022}. While classic formulations consider a single active state driven by positive peer reinforcement, real-world political dynamics in large polarized electorates ($N \sim 10^6 - 10^8$) are inherently bipartisan, involving two competing choices, Candidate $A$ and Candidate $B$. Such collective systems are governed by two competing mechanisms acting on the fractions of individuals committed to each candidate, $\rho_A$ and $\rho_B$: social validation (conformism)~\cite{noelleneumann1974,gerber2008}, wherein citizens require a critical mass of active co-partisans before openly committing to Candidate $A$ or $B$, and cross-candidate rejection (anti-voting)~\cite{medeiros2014,abramowitz2016}, wherein the growth of an opposing candidate instills threat, compelling neutral individuals to mobilize not out of ideological alignment, but to block the opponent. In microscopic social dynamics, local majority-rule and heterogeneous conformist updating have been shown to fundamentally shape collective state transitions, such as enhancing network reciprocity in evolutionary social dilemmas~\cite{szolnoki2015conformity,xia2023reputation}. Here, rather than simulating local microscopic interactions on networks, our formulation abstracts these mechanisms into an aggregate threshold activation response driven by candidate support.

In Granovetter's classic framework~\cite{granovetter1978}, individual decision-making reflects a complex cost-benefit assessment of joining a collective movement: individuals mobilize whenever the perceived net incentive exceeds their intrinsic activation threshold. Here, we extend this threshold concept to a bipartisan setting by modeling the net mobilization incentive for each candidate as a linear combination of positive conformist peer validation and negative anti-voting threat, given by $U_A = \rho_A + \gamma_B \rho_B$ and $U_B = \rho_B + \gamma_A \rho_A$. In this formulation, $\gamma_K \ge 0$ represents the cross-candidate rejection parameter targeted against Candidate $K \in \{A, B\}$: a larger fraction of active partisans $\rho_K$ instills fear in uncommitted voters, generating an anti-$K$ backlash that inflates the effective mobilization drive of the opposing candidate. To represent competing partisan preferences along a unified spectrum, each individual $i=1, \ldots, N$ is assigned a continuous latent ideological coordinate $u_i \in [0, 1]$. Their intrinsic activation thresholds (susceptibilities) for Candidates $A$ and $B$ are defined as $x_i^A = u_i^{1/\lambda_A}$ and $x_i^B = (1 - u_i)^{1/\lambda_B}$, playing the exact same role as the threshold in Granovetter's model: individual $i$ mobilizes for Candidate $A$ if $x_i^A \le U_A$ (or for Candidate $B$ if $x_i^B \le U_B$). The inverse coupling between $u_i$ and $1 - u_i$ captures the anti-correlated nature of partisan preferences, ensuring that a high susceptibility (low threshold) for Candidate $A$ inherently implies a low susceptibility (high threshold) for Candidate $B$.

The mobilization readiness of each candidate's political base is governed by candidate-specific cumulative threshold distributions $F_A(x) = x^{\lambda_A}$ and $F_B(y) = y^{\lambda_B}$. Sociologically, the shape exponents $\lambda_A, \lambda_B > 0$ parametrize the intrinsic activation barriers of the respective electorates: $\lambda_K < 1$ represents a susceptible, self-starting base for Candidate $K \in \{A, B\}$ whose voters mobilize readily under minimal social pressure; conversely, $\lambda_K > 1$ models a hesitant base with high activation barriers, where voters require substantial social validation or perceived threat to participate. In large electorates where finite-size fluctuations are negligible ($1/\sqrt{N} \to 0$), this minimal formulation yields an exact, deterministic two-dimensional piecewise map.

In this Letter, we map the complete phase portrait of this bipartisan map and establish exact analytical criteria for its dynamic instabilities. Focusing on symmetric electorates ($\lambda_A = \lambda_B = \lambda$ and $\gamma_A = \gamma_B = \gamma$), we demonstrate that cross-candidate rejection reorganizes the phase space between an under-mobilized regime (where a fraction $\rho_0 > 0$ abstains) and a fully mobilized regime (where $\rho_0 = 0$). We uncover three distinct bifurcation mechanisms: (i) spontaneous symmetry breaking into candidate-dominant lead lock-in states via a pitchfork bifurcation ($\gamma_{\text{pf}}$) intrinsic to radicalized bases ($\lambda < 1/2$); (ii) a non-smooth border-collision bifurcation ($\gamma_{\text{bound}}$) governing sudden transitions to complete abstention ($\rho_0 = 1$) in hesitant electorates ($\lambda > 1$); and (iii) a period-doubling flip bifurcation ($\gamma_{\text{flip}}$) driving persistent period-2 anti-frontrunner polling swings, which exhibits a subcritical flip wedge ($2 < \gamma < \gamma_{\text{flip}}$) for moderately radicalized electorates ($1/2 < \lambda < 1$). Interpreting the discrete step $t \to t+1$ as macro campaign evaluation rounds, we demonstrate how negative partisanship turns campaign timing and initial perceptions into critical determinants of election outcomes.

\section{Exact Piecewise Discrete Map}

Rather than tracking inter-electoral cycles across successive years, the discrete time index $t = 0, 1, 2, \dots$ in our formulation parametrizes the intra-campaign mobilization dynamics within a single election cycle. Physically, each step $t \to t+1$ represents an opinion-updating round---such as the release of sequential public polling reports or major campaign events---where voters reassess candidate viability, social momentum, and perceived threats.

At campaign step $t$, the state of the electorate is defined by the vector of voter intentions $\boldsymbol{\rho}^{(t)} = (\rho_A^{(t)}, \rho_B^{(t)}, \rho_0^{(t)})$, subject to the conservation condition $\rho_A^{(t)} + \rho_B^{(t)} + \rho_0^{(t)} = 1$, where $\rho_0^{(t)}$ denotes the uncommitted fraction. Crucially, $\rho_0^{(t)} = 1 - \rho_A^{(t)} - \rho_B^{(t)}$ parametrizes politically inactive or undecided individuals at campaign round $t$. During intra-campaign polling rounds, $\rho_0^{(t)}$ reflects temporary indecision or non-commitment, whereas on election day ($t_{\text{elect}}$), $\rho_0^{(t_{\text{elect}})}$ represents final electoral abstention (non-voting turnout). Throughout this Letter, we maintain this distinction, reserving ``undecided'' for campaign-stage survey rounds and ``abstention'' for final electoral outcomes. Upon observing the political landscape at step $t$, the effective mobilization drives perceived by the public are formulated as
\begin{align}
U_A &= \rho_A^{(t)} + \gamma_B \rho_B^{(t)}, \label{eq:UA} \\
U_B &= \rho_B^{(t)} + \gamma_A \rho_A^{(t)}, \label{eq:UB}
\end{align}
where $\gamma_K \ge 0$ measures the degree to which support for Candidate $K=\{A,B\}$ induces cross-candidate rejection or defensive mobilization in the opposing camp.

To derive the updated voter fractions at step $t+1$, we identify the regions of the latent ideological coordinate $u \in [0, 1]$ satisfying the activation criteria $x^A(u) \le U_A$ and $x^B(u) \le U_B$. Because $u$ is uniformly distributed, the activation conditions $u^{1/\lambda_A} \le U_A$ for Candidate $A$ and $(1 - u)^{1/\lambda_B} \le U_B$ for Candidate $B$ define two mobilization intervals in $u$-space
\begin{equation}
u \in [0, U_A^{\lambda_A}] \qquad \text{and} \qquad u \in [1 - U_B^{\lambda_B}, 1].
\label{eq:activation_intervals}
\end{equation}
The relative overlap or separation of these two intervals determines the updated distribution of voter intentions $\boldsymbol{\rho}^{(t+1)}$ in the subsequent campaign round.
\subsection{Non-Overlapping Activation}
When $U_A^{\lambda_A} + U_B^{\lambda_B} < 1$, the activation intervals $[0, U_A^{\lambda_A}]$ and $[1 - U_B^{\lambda_B}, 1]$ are disjoint. Integration over these independent regions yields
\begin{align}
\rho_A^{(t+1)} &= \int_0^{U_A^{\lambda_A}} du = U_A^{\lambda_A} = (\rho_A^{(t)} + \gamma_B \rho_B^{(t)})^{\lambda_A}, \label{eq:mapA_I} \\
\rho_B^{(t+1)} &= \int_{1 - U_B^{\lambda_B}}^1 du = U_B^{\lambda_B} = (\rho_B^{(t)} + \gamma_A \rho_A^{(t)})^{\lambda_B}. \label{eq:mapB_I}
\end{align}
Physically, this reflects the core threshold principle: the fraction of individuals mobilizing for Candidate $K=\{A,B\}$ at step $t+1$ corresponds precisely to the mass of voters whose intrinsic activation thresholds $x^K(u)$ lie below the effective drive $U_K$.
The uncommitted population in the central buffer zone forms the non-zero abstention fraction
\begin{equation}
\rho_0^{(t+1)} = 1 - \rho_A^{(t+1)} - \rho_B^{(t+1)} = 1 - U_A^{\lambda_A} - U_B^{\lambda_B} > 0.
\label{eq:rho0_I}
\end{equation}

\subsection{Overlapping Activation}
When $U_A^{\lambda_A} + U_B^{\lambda_B} \ge 1$, the activation intervals overlap in $u$-space. Every individual meets the activation threshold for at least one candidate, driving abstention to complete collapse, $\rho_0^{(t+1)} = 0$. In the overlapping region, individuals qualify for both candidates and break the tie by choosing the candidate offering the higher net conviction surplus, $S_K(u) = U_K - x^K(u)$. The decision boundary $u_{\text{cut}}^{(t)} \in [0, 1]$ separating Candidate $A$ voters from Candidate $B$ voters is the unique root satisfying equal conviction surplus $S_A(u_{\text{cut}}) = S_B(u_{\text{cut}})$,
\begin{equation}
U_A - (u_{\text{cut}}^{(t)})^{1/\lambda_A} = U_B - (1 - u_{\text{cut}}^{(t)})^{1/\lambda_B}.
\label{eq:ucut}
\end{equation}
Voters with $u \le u_{\text{cut}}^{(t)}$ choose Candidate $A$, while those with $u > u_{\text{cut}}^{(t)}$ choose Candidate $B$. Integrating over these regions yields
\begin{align}
\rho_A^{(t+1)} &= \int_0^{u_{\text{cut}}^{(t)}} du = u_{\text{cut}}^{(t)}, \label{eq:mapA_II} \\
\rho_B^{(t+1)} &= \int_{u_{\text{cut}}^{(t)}}^1 du = 1 - u_{\text{cut}}^{(t)}. \label{eq:mapB_II}
\end{align}

\subsection{Unified Piecewise Discrete Map}
The dynamic condition $U_A^{\lambda_A} + U_B^{\lambda_B} = 1$ naturally defines a switching boundary  in the phase portrait $(\rho_A, \rho_B)$, partitioning the phase space into a Moderate Domain $\mathcal{D}_{\text{I}}$ ($\rho_0 > 0$) and a Hyper-Polarized Domain $\mathcal{D}_{\text{II}}$ ($\rho_0 = 0$). The full 2D deterministic discrete map $(\rho_A^{(t)}, \rho_B^{(t)}) \mapsto (\rho_A^{(t+1)}, \rho_B^{(t+1)})$ is explicitly given by
\begin{eqnarray}\label{eq:full_mapI}
\rho_A^{(t+1)} & = &   (\rho_A^{(t)} + \gamma_B \rho_B^{(t)})^{\lambda_A} \\
 \rho_B^{(t+1)} & = & (\rho_B^{(t)} + \gamma_A \rho_A^{(t)})^{\lambda_B} 
\end{eqnarray}
if $U_A^{\lambda_A} + U_B^{\lambda_B} < 1$ (Domain  $\mathcal{D}_{\text{I}}$),  and
\begin{eqnarray}\label{eq:full_mapII}
\rho_A^{(t+1)} & = &   u_{\text{cut}}^{(t)}  \\
 \rho_B^{(t+1)} & = &1 - u_{\text{cut}}^{(t)} 
\end{eqnarray}
if  $U_A^{\lambda_A} + U_B^{\lambda_B} \ge 1$ (Domain  $\mathcal{D}_{\text{II}}$).

While asymmetric parameters ($\lambda_A \neq \lambda_B, \gamma_A \neq \gamma_B$) allow the model to describe unbalanced political landscapes---such as an enthusiastic, self-mobilizing movement competing against a hesitant electorate---we focus henceforth on the fully symmetric regime ($\lambda_A = \lambda_B = \lambda$ and $\gamma_A = \gamma_B = \gamma$). This baseline effectively models deeply polarized bipartisan contests, such as the 2022 and 2026 Brazilian presidential elections, where opposing candidates command symmetrically high rejection rates and highly radicalized bases, resulting in an almost exact $50/50$ split of the active electorate~\cite{samuels2024,petherick2026}.

\section{Fixed Point and Stability Analysis}

Under parameter symmetry ($\lambda_A = \lambda_B = \lambda$ and $\gamma_A = \gamma_B = \gamma$), the 2D piecewise discrete map simplifies to
\begin{equation}
\begin{pmatrix} \rho_A^{(t+1)} \\[6pt] \rho_B^{(t+1)} \end{pmatrix} = 
\begin{cases}
\begin{pmatrix} (\rho_A^{(t)} + \gamma \rho_B^{(t)})^{\lambda} \\[6pt] (\rho_B^{(t)} + \gamma \rho_A^{(t)})^{\lambda} \end{pmatrix}, & \text{if } U_A^\lambda + U_B^\lambda < 1 \quad (\text{Domain } \mathcal{D}_{\text{I}}), \\[14pt]
\begin{pmatrix} u_{\text{cut}}^{(t)} \\[6pt] 1 - u_{\text{cut}}^{(t)} \end{pmatrix}, & \text{if } U_A^\lambda + U_B^\lambda \ge 1 \quad (\text{Domain } \mathcal{D}_{\text{II}}),
\end{cases}
\label{eq:sym_map}
\end{equation}
where in Domain $\mathcal{D}_{\text{II}}$, $u_{\text{cut}}^{(t)}$ is defined by the surplus condition $(u_{\text{cut}}^{(t)})^{1/\lambda} - (1 - u_{\text{cut}}^{(t)})^{1/\lambda} = U_A - U_B$. For the linear threshold case ($\lambda = 1$), this yields  $u_{\text{cut}}^{(t)} = \gamma/2 + (1 - \gamma)\rho_A^{(t)}$ on the fully mobilized manifold ($\rho_A^{(t)} + \rho_B^{(t)} = 1$). The stationary states $\boldsymbol{\rho}^* = (\rho_A^*, \rho_B^*, \rho_0^*)$ and their local stability are obtained by linearizing Eq.~\eqref{eq:sym_map} around each fixed point~\cite{strogatz2015}.

\subsection{Solutions in Domain $\mathcal{D}_{\text{I}}$}

Domain $\mathcal{D}_{\text{I}}$ represents the under-mobilized regime where voter turnout remains incomplete ($\rho_0 > 0$). Setting $\rho_A^* = \rho_B^* = 0$ yields the pure apathy state $\boldsymbol{\rho}^0 = (0, 0, 1)$. Linearization around $\boldsymbol{\rho}^0$ shows that its Jacobian eigenvalues vanish ($\mu_{1,2} = 0$), rendering it a super-stable sink for hesitant electorates ($\lambda > 1$). 
For $\lambda = 1$, the eigenvalues simplify to $\mu_1 = 1 + \gamma$ and $\mu_2 = 1 - \gamma$.  Consequently, the apathy state $\boldsymbol{\rho}^0$ is unstable for all $\gamma > 0$. In the special degenerate case $(\lambda, \gamma) = (1, 0)$, both eigenvalues equal $1$ ($\mu_1 = \mu_2 = 1$), reducing the mapping in Domain $\mathcal{D}_{\text{I}}$ to the identity where every state is a neutrally stable fixed point.  For $0 < \lambda < 1$, the derivatives of the map diverge at the origin, reflecting the absence of an activation threshold barrier for self-starting electorates, which makes $\boldsymbol{\rho}^0$ an unstable repellor.

Non-trivial active fixed points ($\rho_A^*, \rho_B^* > 0$) within Domain $\mathcal{D}_{\text{I}}$ exist only for hesitant electorates ($\lambda > 1$), taking the symmetric form $\rho_A^* = \rho_B^* = (1 + \gamma)^{\lambda/(1 - \lambda)}$, with abstention fraction $\rho_0^* = 1 - 2 (1 + \gamma)^{\lambda/(1 - \lambda)}$. Requiring $\rho_0^* > 0$ defines a critical threshold
\begin{equation}
\gamma > \gamma_{\text{bound}}(\lambda) \equiv 2^{1 - 1/\lambda} - 1.
\label{eq:gamma_bound_cond}
\end{equation}
Geometrically, the boundary separating under-mobilized dynamics (Domain $\mathcal{D}_{\text{I}}$) from full mobilization (Domain $\mathcal{D}_{\text{II}}$) is the switching surface $U_A^\lambda + U_B^\lambda = 1$ ($\rho_0 = 0$).  As cross-rejection decreases toward $\gamma_{\text{bound}}(\lambda)$, the abstention fraction shrinks ($\rho_0^* \to 0$) until the active state hits this switching boundary. In piecewise-smooth dynamical systems, the annihilation or structural change of a fixed point upon reaching a switching surface is termed a border-collision bifurcation~\cite{dibernardo2008}.

When $\gamma > \gamma_{\text{bound}}(\lambda)$, linear stability analysis confirms that this active state $\boldsymbol{\rho}^{*,\text{sad}}$ is a saddle point, with leading eigenvalue $\mu_1 = \lambda > 1$ along the line of symmetry ($\rho_A = \rho_B$) and transverse eigenvalue $\mu_2 = \lambda (1 - \gamma) / (1 + \gamma)$. The stable manifold of $\boldsymbol{\rho}^{*,\text{sad}}$ acts as a phase-space separatrix between mobilizations decaying to pure apathy $\boldsymbol{\rho}^0$ and those crossing the domain border into Domain $\mathcal{D}_{\text{II}}$~\cite{dibernardo2008}. At $\gamma = \gamma_{\text{bound}}(\lambda)$, this saddle point collides with the switching boundary and disappears; for $\gamma < \gamma_{\text{bound}}(\lambda)$, it no longer exists in Domain $\mathcal{D}_{\text{I}}$, leaving $\boldsymbol{\rho}^0$ as the sole attractor across the domain.

\subsection{Solutions in Domain $\mathcal{D}_{\text{II}}$}

In Domain $\mathcal{D}_{\text{II}}$, voter abstention completely vanishes ($\rho_0^* = 0$), collapsing the dynamics to the 1D mapping 
\begin{equation}
\rho_A^{(t+1)} = u_{\text{cut}}(\rho_A^{(t)}), 
\label{eq:mapII}
\end{equation}
 whose stationary states satisfy
\begin{equation}\label{eq:fp_II_sym}
(\rho_A^*)^{1/\lambda} - (1 - \gamma) \rho_A^* = (1 - \rho_A^*)^{1/\lambda} - (1 - \gamma) (1 - \rho_A^*).
\end{equation}
The symmetric share $\rho_A^* = 1/2$ is an exact solution to Eq.~\eqref{eq:fp_II_sym}, corresponding to the balanced state $\boldsymbol{\rho}^* = (1/2, 1/2, 0)$. Requiring $\boldsymbol{\rho}^*$ to satisfy the Domain $\mathcal{D}_{\text{II}}$ existence condition $U_A^\lambda + U_B^\lambda \ge 1$ defines the lower existence boundary $\gamma \ge \gamma_{\text{bound}}(\lambda) \equiv 2^{1 - 1/\lambda} - 1$, which is positive for hesitant electorates ($\lambda > 1$) and trivially satisfied ($\gamma_{\text{bound}} \le 0$) for all physical $\gamma \ge 0$ when $\lambda \le 1$. Linearization of Eq.~\eqref{eq:fp_II_sym} around $\rho_A^* = 1/2$ yields the 1D characteristic eigenvalue
\begin{equation}
\mu = \left. \frac{d u_{\text{cut}}}{d \rho_A} \right|_{\rho_A = 1/2} = f(\lambda)(1 - \gamma), 
\label{eq:eigenvalue}
\end{equation}
where $f(\lambda) \equiv \lambda 2^{1/\lambda - 1}$. The symmetric state $\boldsymbol{\rho}^*$ is locally stable provided $|\mu| < 1$. Its loss of stability is governed by smooth bifurcations inside Domain $\mathcal{D}_{\text{II}}$ as well as non-smooth border-collision bifurcations at the switching boundary~\cite{dibernardo2008}.

For highly radicalized voter bases ($\lambda < 1/2$), $f(\lambda) > 1$ while $\gamma_{\text{bound}} = 0$. Here, a supercritical pitchfork bifurcation ($\mu = +1$) occurs inside Domain $\mathcal{D}_{\text{II}}$ when cross-rejection is low ($\gamma < \gamma_{\text{pf}} = 1 - 1/f(\lambda)$). Below $\gamma_{\text{pf}}$, $\boldsymbol{\rho}^*$ loses stability to a pair of stable asymmetric fixed points $\boldsymbol{\rho}^{*, A} = (\rho_*, 1 - \rho_*, 0)$ and $\boldsymbol{\rho}^{*, B} = (1 - \rho_*, \rho_*, 0)$ with $\rho_* > 1/2$ (Asymmetric Phase). This represents spontaneous symmetry breaking: initial campaign perceptions $\boldsymbol{\rho}^{(0)}$ select which candidate locks in a decisive polling lead prior to election day. Conversely, for moderately radicalized electorates ($1/2 \le \lambda \le 1$), $f(\lambda) \le 1$, rendering $\boldsymbol{\rho}^*$ immune to pitchfork instabilities for all $\gamma \ge 0$.

For hesitant electorates ($\lambda > 1$), formal evaluation of $\gamma_{\text{pf}}(\lambda) = 1 - 1/f(\lambda)$ yields a nominal threshold, but $\gamma_{\text{pf}}(\lambda) < \gamma_{\text{bound}}(\lambda)$ holds across the entire range $\lambda > 1$. As cross-rejection $\gamma$ is lowered, $\boldsymbol{\rho}^*$ does not undergo a smooth pitchfork bifurcation inside Domain $\mathcal{D}_{\text{II}}$. Instead, at $\gamma = \gamma_{\text{bound}}(\lambda)$, $\boldsymbol{\rho}^*$ collides with the switching boundary $U_A^\lambda + U_B^\lambda = 1$ and vanishes via a non-smooth border-collision bifurcation~\cite{dibernardo2008}. Below $\gamma_{\text{bound}}(\lambda)$, candidate-dominant solutions predicted by Eq.~\eqref{eq:fp_II_sym} are virtual fixed points residing in Domain $\mathcal{D}_{\text{I}}$ ($U_A^\lambda + U_B^\lambda < 1$), rendering them physically inaccessible. In this regime, the super-stable apathy sink $\boldsymbol{\rho}^0$ becomes the sole attractor, whose basin of attraction encompasses the entire phase space.

Finally, for any shape exponent $\lambda > 0$, elevated cross-rejection induces a period-doubling flip bifurcation ($\mu = -1$) when $\gamma$ exceeds $\gamma_{\text{flip}}(\lambda) \equiv 1 + 1/f(\lambda)$. Exceeding $\gamma_{\text{flip}}$ destabilizes the symmetric equilibrium $\boldsymbol{\rho}^*$, governing the onset of Period-2 anti-frontrunner polling swings $\{\boldsymbol{\rho}^{(1)}, \boldsymbol{\rho}^{(2)}\}$, where $\boldsymbol{\rho}^{(1)} = (r, 1 - r, 0)$ and $\boldsymbol{\rho}^{(2)} = (1 - r, r, 0)$ with $r \in (1/2, 1]$. Applying the Period-2 orbit condition $\rho_A^{(t+1)}=r$ and $u_{\text{cut}}(\rho_A^{(t)})= 1 - r$ to Eq.~\eqref{eq:mapII} yields the exact implicit equation governing the oscillation amplitude $r$ as a function of cross-rejection,
\begin{equation}
\gamma(r) = 1 + \frac{r^{1/\lambda} - (1 - r)^{1/\lambda}}{2r - 1},
\label{eq:period2_amplitude}
\end{equation}
where $r \to 1/2$ recovers $\gamma_{\text{flip}}$ and $r \to 1$ yields complete polarization at $\gamma = 2$. The local stability of this cycle is governed by the second-iterate multiplier $\Lambda_2 = [2(1 - \gamma) / (a [r^{a-1} + (1 - r)^{a-1}])]^2$ with $a = 1/\lambda$. For $\lambda < 1/2$ ($a > 2$) and $\lambda > 1$ ($a < 1$), $\Lambda_2 < 1$, rendering the flip supercritical with a stable Period-2 cycle growing continuously from $r = 1/2$ at $\gamma_{\text{flip}}$ to $r = 1$ at $\gamma = 2$. Conversely, for $1/2 < \lambda < 1$ ($1 < a < 2$), $\Lambda_2 > 1$ near $r = 1/2$, making the flip subcritical: the emerging Period-2 orbit is unstable, while the fully polarized cycle $(1,0) \leftrightarrow (0,1)$ exists and is stable for all $\gamma > 2$, creating a bistable wedge for $2 < \gamma < \gamma_{\text{flip}}$ where the stable symmetric equilibrium coexists with the fully polarized cycle. At the degenerate boundaries $\lambda = 1/2$ and $\lambda = 1$, $\gamma_{\text{flip}} = 2$ and the map degenerates into an exact reflection. Here, intense negative partisanship transforms campaign dynamics into a rhythmic anti-frontrunner rejection engine, where a candidate lead surge in one evaluation round triggers an immediate opposing backlash in the next. We note that our piecewise map assumes instantaneous preference updates at the zero-surplus threshold $S_A(u) - S_B(u) = 0$. Introducing voter switching friction---such as psychological partisan inertia~\cite{taber2006motivated} or cognitive switching costs~\cite{martins2008continuous}---would incorporate micro-hysteresis into opinion updates, thereby shifting the period-doubling threshold $\gamma_{\text{flip}}$ toward higher cross-rejection values and dampening campaign oscillation amplitudes.

\begin{figure*}[t!]
\centering
\includegraphics[width=\linewidth]{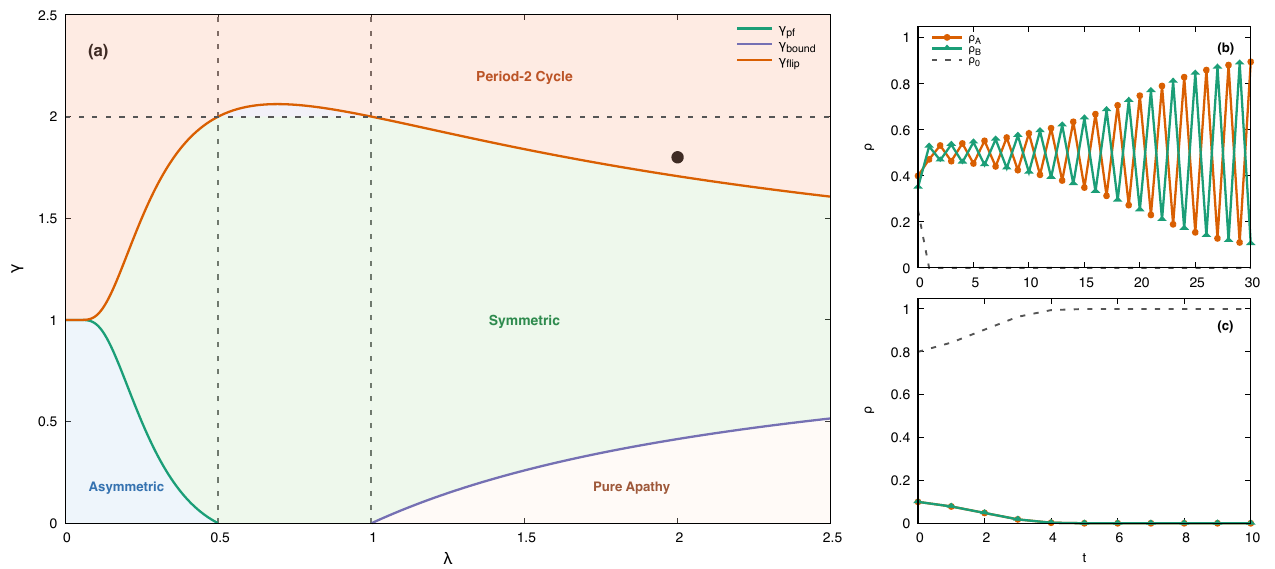}
\caption{(Color online) Dynamical regimes and initial-condition sensitivity in intra-campaign opinion dynamics. (a) Phase diagram in the $(\lambda, \gamma)$ parameter plane. Solid curves denote the pitchfork ($\gamma_{\text{pf}}$), border-collision ($\gamma_{\text{bound}}$), and flip ($\gamma_{\text{flip}}$) bifurcations. The vertical dashed line $\lambda = 1/2$ bounds the Asymmetric Phase ($\lambda < 1/2$), while $\lambda = 1$ separates self-starting voter bases ($\lambda < 1$) from hesitant voter bases ($\lambda > 1$). For hesitant electorates ($\lambda > 1$), candidate-dominant fixed points in Domain $\mathcal{D}_{\text{II}}$ are virtual; below $\gamma_{\text{bound}}(\lambda)$, all initial states decay to the super-stable apathy sink $\boldsymbol{\rho}^0 = (0,0,1)$ (Pure Apathy domain).  For $1/2 < \lambda < 1$, the region $2 < \gamma < \gamma_{\text{flip}}$ exhibits subcritical bistability between the symmetric state and the fully polarized Period-2 cycle. The horizontal dashed line $\gamma = 2$ marks the onset of complete cycle saturation ($100\%$ turnout). The black dot indicates the representative operating point $(\lambda = 2, \gamma = 1.8)$. (b--c) Intra-campaign polling trajectory evolution of candidate shares $\rho_A(t)$ (orange), $\rho_B(t)$ (teal), and undecided fraction $\rho_0(t)$ (dashed grey) across survey rounds $t$ at $(\lambda = 2, \gamma = 1.8)$. (b) Above-threshold initial mobilization ($\rho_0^{(0)} = 0.25$) drives complete undecided collapse ($\rho_0 \to 0$) into a stable Period-2 anti-frontrunner cycle. (c) Sub-threshold initial mobilization ($\boldsymbol{\rho}^{(0)} = (0.1, 0.1, 0.8)$) decays monotonically into the super-stable apathy sink $\boldsymbol{\rho}^0 = (0,0,1)$.}
\label{fig:phase_diagram}
\end{figure*}

The dynamic structure described by these bifurcation boundaries is exhibited in Fig.~\ref{fig:phase_diagram}(a). In the low-rejection regime ($\gamma < \gamma_{\text{pf}}$), asymmetric polling equilibria exist only where $f(\lambda) > 1$, restricting the Asymmetric Phase to highly radicalized bases ($\lambda < 1/2$). Here, self-starting activation renders the apathy state $\boldsymbol{\rho}^0 = (0,0,1)$ unstable, dividing the phase space into two mirror-symmetric basins of attraction that determine which candidate-dominant fixed point ($\boldsymbol{\rho}^{*, A}$ or $\boldsymbol{\rho}^{*, B}$) establishes a decisive campaign lead. For moderately radicalized electorates ($1/2 \le \lambda \le 1$), $f(\lambda) \le 1$ suppresses pitchfork instabilities for all $\gamma \ge 0$; the apathy state remains unstable, but trajectories converge to the balanced symmetric equilibrium $\boldsymbol{\rho}^* = (1/2, 1/2, 0)$ for $\gamma < \min(2, \gamma_{\text{flip}})$, while for $1/2 < \lambda < 1$ and $2 < \gamma < \gamma_{\text{flip}}$, $\boldsymbol{\rho}^*$ coexists with the fully polarized Period-2 cycle. In contrast, hesitant voter bases ($\lambda > 1$) possess a super-stable complete abstention sink arising from high initial activation barriers. Here, the nominal pitchfork threshold $\gamma_{\text{pf}}$ corresponds to virtual fixed points; instead, the border-collision boundary $\gamma_{\text{bound}}(\lambda) = 2^{1 - 1/\lambda} - 1$ separates a Pure Apathy domain ($\gamma < \gamma_{\text{bound}}$)---where $\boldsymbol{\rho}^0$ is the sole global attractor---from a bistable regime ($\gamma_{\text{bound}} < \gamma < \gamma_{\text{flip}}$) where apathy coexists with the symmetric 50/50 state $\boldsymbol{\rho}^*$.

For higher cross-candidate rejection ($\gamma > \gamma_{\text{flip}}$), this phase-space partition governs the representative operating point $(\lambda = 2, \gamma = 1.8)$. As shown in Fig.~\ref{fig:phase_diagram}(c), if initial mobilization falls below the saddle-separatrix threshold in Domain $\mathcal{D}_{\text{I}}$,  campaign engagement monotonically decays into complete voter burnout at $\boldsymbol{\rho}^0$. Conversely, an above-threshold initial state ($\rho_0^{(0)} = 0.25$, Fig.~\ref{fig:phase_diagram}(b)) amplifies engagement until total mobilization is reached ($\rho_0 \to 0$), whereupon the system enters Domain $\mathcal{D}_{\text{II}}$ and settles into persistent anti-front\-run\-ner polling swings. Remarkably, if the system is initialized with perfectly balanced turnout ($\rho_A^{(0)} = \rho_B^{(0)} = 0.2$), the dynamics evolve strictly along the 1D invariant manifold of political symmetry ($\rho_A = \rho_B$). In this subspace, mobilization expands through Domain $\mathcal{D}_{\text{I}}$ until landing directly on the balanced fixed point $\boldsymbol{\rho}^* = (1/2, 1/2, 0)$ in Domain $\mathcal{D}_{\text{II}}$. However, because $\gamma = 1.8 > \gamma_{\text{flip}}(2.0) \approx 1.71$ places the system past the period-doubling threshold, $\boldsymbol{\rho}^*$ is transversely unstable with eigenvalue $\mu = f(2.0)(1 - 1.8) \approx -1.13$. Consequently, any infinitesimal symmetry-breaking perturbation ($\rho_A \neq \rho_B$) immediately ejects the trajectory off this invariant subspace onto the robust period-2 polling attractor in Fig.~\ref{fig:phase_diagram}(b).

\begin{figure*}[t!]
\centering
\includegraphics[width=\linewidth]{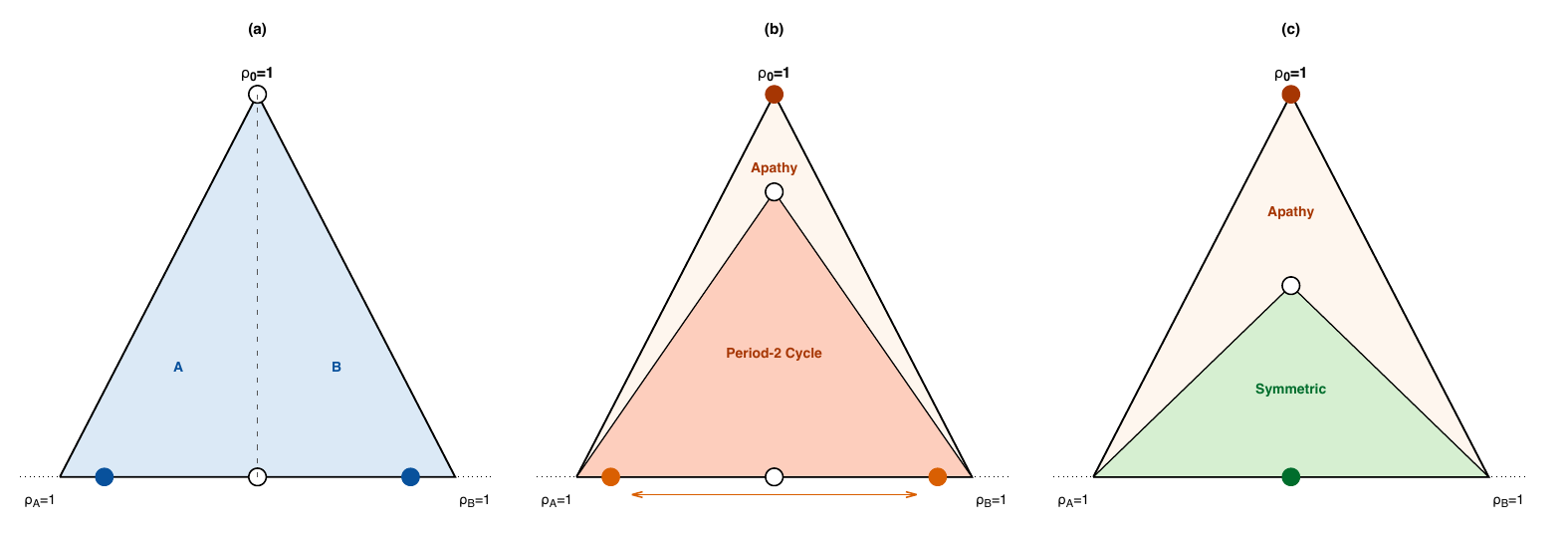}
\caption{(Color online) Basins of attraction on the electoral simplex.
(a) $\lambda=0.25, \gamma=0.2$ (Asymmetric Phase): Highly radicalized electorates ($\lambda < 1/2$) undergo spontaneous symmetry breaking via a pitchfork bifurcation, partitioning the phase space into two mirror-symmetric basins leading to candidate lead lock-in ($\boldsymbol{\rho}^{*,A}$ or $\boldsymbol{\rho}^{*,B}$).
(b) $\lambda=2, \gamma=1.8$ (Period-2 Cycle): Elevated cross-candidate rejection ($\gamma > \gamma_{\text{flip}} \approx 1.71$) destabilizes the symmetric state $\boldsymbol{\rho}^* = (1/2, 1/2, 0)$ (empty circle at base) into a period-2 polling cycle oscillating between $\boldsymbol{\rho}^{(1)}$ and $\boldsymbol{\rho}^{(2)}$ (filled orange circles connected by a double-headed arrow). The stable manifold of the Domain $\mathcal{D}_{\text{I}}$ saddle point (empty circle at height $\rho_0 \approx 0.74$) partitions the simplex into the Apathy Basin and the Period-2 Cycle Basin.
(c) $\lambda=2, \gamma=1$ (Symmetric Phase): Moderate cross-rejection ($\gamma_{\text{bound}} < \gamma < \gamma_{\text{flip}}$) yields bistability between apathy $\boldsymbol{\rho}^0$ (top filled orange circle) and the symmetric 50/50 state $\boldsymbol{\rho}^* = (1/2, 1/2, 0)$ (filled green circle at base), separated by the stable manifold of the interior Domain $\mathcal{D}_{\text{I}}$ saddle point (empty circle at height $\rho_0 = 0.5$).}
\label{fig:simplexes}
\end{figure*}

The basin-of-attraction organization underlying these dynamical transitions is visualized on the electoral simplex in Fig.~\ref{fig:simplexes}. In the Asymmetric Phase ($\lambda = 0.25, \gamma = 0.2$, Fig.~\ref{fig:simplexes}(a)), highly radicalized electorates ($\lambda < 1/2$) exhibit global self-starting activation. Because the apathy state $\boldsymbol{\rho}^0 = (0,0,1)$ at the top vertex is unstable (a repellor/saddle), any non-zero initial mobilization is amplified away from total abstention. The phase space is neatly partitioned into two symmetric basins of attraction by the central invariant line of symmetry $\rho_A = \rho_B$, dictating convergence to either the Candidate A polling dominance sink $\boldsymbol{\rho}^{*,A}$ or the Candidate B polling dominance sink $\boldsymbol{\rho}^{*,B}$. Notably, along the simplex edges where one candidate has zero initial support (e.g., $\rho_A = 0$), cross-candidate rejection ($\gamma > 0$) acts as an automatic recruitment engine: undecided voters experiencing negative sentiment against Candidate B ($U_A = \gamma \rho_B > 0$) register protest support for Candidate A in opinion surveys, pulling trajectories off the boundary into the interior.

In the period-2 cycle regime ($\lambda = 2, \gamma = 1.8$, Fig.~\ref{fig:simplexes}(b)), hesitant voter activation ($\lambda > 1$) turns the apathy vertex $\boldsymbol{\rho}^0$ into a super-stable sink, while elevated cross-candidate rejection ($\gamma > \gamma_{\text{flip}} \approx 1.71$) destabilizes the balanced state $\boldsymbol{\rho}^* = (1/2, 1/2, 0)$ at the base midpoint via a period-doubling flip bifurcation. The stable manifold of the interior Domain $\mathcal{D}_{\text{I}}$ saddle point (located at abstention fraction $\rho_0^* \approx 0.74$) partitions the simplex into two primary basins: an upper Apathy Basin where sub-separatrix initial mobilization decays into complete voter burnout, and a lower Period-2 Cycle Basin where sufficiently mobilized initial states amplify until full turnout is reached ($\rho_0 \to 0$), settling into persistent anti-frontrunner polling swings between $\boldsymbol{\rho}^{(1)}$ and $\boldsymbol{\rho}^{(2)}$. Finally, for moderate cross-candidate rejection ($\lambda = 2, \gamma = 1$, Fig.~\ref{fig:simplexes}(c)), the system enters the Symmetric Phase ($\gamma_{\text{bound}} < \gamma < \gamma_{\text{flip}}$). Here, the state $\boldsymbol{\rho}^* = (1/2, 1/2, 0)$ becomes locally stable, yielding a bistable phase space where the stable manifold of the Domain $\mathcal{D}_{\text{I}}$ saddle point (at height $\rho_0^* = 0.5$) separates the apathy basin from the symmetric 50/50 equilibrium basin.

\section{Conclusions}

In the broader context of sociophysics, models pioneered by Galam~\cite{galam2008,galam2018} have successfully captured voting anomalies and turnout fluctuations through local majority-rule dynamics. Departing from local spatial update rules, our continuous two-dimensional piecewise map captures aggregate campaign dynamics where non-voting is an endogenous phase variable originating from a uniform latent ideological spectrum $u \in [0, 1]$. By examining a fully symmetric scenario ($\lambda_A = \lambda_B = \lambda$ and $\gamma_A = \gamma_B = \gamma$), we demonstrate that cross-candidate rejection drives a sharp transition from the under-mobilized regime ($\mathcal{D}_{\text{I}}$) to the fully mobilized regime ($\mathcal{D}_{\text{II}}$), organizing the phase space through non-smooth tipping dynamics.

Despite complete candidate parity, the system exhibits rich multi-stability. For highly radicalized electorates ($\lambda < 1/2$), strong self-starting voter activation spontaneously breaks political symmetry at low rejection levels ($\gamma < \gamma_{\text{pf}}$), partitioning the phase space into two mirror-symmetric candidate-dominant basins. Increasing cross-candidate rejection $\gamma$ counteracts this intrinsic radicalization, eventually merging the asymmetric states back into political symmetry at $\gamma = \gamma_{\text{pf}}$ via a pitchfork bifurcation. Conversely, for hesitant electorates ($\lambda > 1$), a non-smooth border-collision bifurcation at $\gamma_{\text{bound}}(\lambda)$ opens a bistable regime for $\gamma > \gamma_{\text{bound}}(\lambda)$, where complete abstention $\boldsymbol{\rho}^0$ coexists with a mobilized equilibrium. In this regime, $\boldsymbol{\rho}^0$ acts as a super-stable sink whenever initial mobilization falls below the saddle separatrix. Across these multi-stable regions, the initial condition vector $\boldsymbol{\rho}^{(0)} = (\rho_A^{(0)}, \rho_B^{(0)}, \rho_0^{(0)})$---parametrizing early public perceptions, media framing, or candidate rollouts at campaign onset---acts as a deterministic catalyst selecting the ultimate campaign trajectory.

For elevated cross-candidate rejection ($\gamma > \gamma_{\text{flip}}$), the symmetric equilibrium loses stability via a period-doubling flip bifurcation, driving the system into a period-2 cycle of anti-frontrunner polling swings. While this transition is supercritical for $\lambda < 1/2$ and $\lambda > 1$, it is subcritical for moderately radicalized electorates ($1/2 < \lambda < 1$), creating a bistable wedge for $2 < \gamma < \gamma_{\text{flip}}$ where the balanced state coexists with a fully polarized cycle. Operating on discrete macro evaluation rounds ($t \to t+1$), this dynamic reveals that the final election outcome is fundamentally dictated by the temporal alignment of election day relative to the campaign polling phase. An overwhelming lead in one survey round triggers an intense negative partisan backlash in the next, turning polarized elections into highly time-sensitive dynamical systems where early campaign momentum, voter mobilization, and voting timing interact to shape democratic outcomes.
\section*{Acknowledgments}

JFF is partially supported by Conselho Nacional de Desenvolvimento Cient\'{\i}fico e Tecnol\'ogico grant number 300200/2026-9.

\end{document}